\documentclass[pdflatex,sn-nature]{sn-jnl}

\usepackage{graphicx}%
\usepackage{multirow}%
\usepackage{amsmath,amssymb,amsfonts}%
\usepackage{amsthm}%
\usepackage{mathrsfs}%
\usepackage[title]{appendix}%
\usepackage{xcolor}%
\usepackage{textcomp}%
\usepackage{manyfoot}%
\usepackage{booktabs}%
\usepackage{algorithm}%
\usepackage{algorithmicx}%
\usepackage{algpseudocode}%
\usepackage{listings}%
\usepackage{array}

\usepackage{tabularx}
\usepackage{booktabs}
\usepackage{makecell}  
\usepackage{calc}      
\usepackage[english]{babel}
\usepackage{csquotes}

\newcommand{\Nino}{Ni\~no}

\theoremstyle{thmstyleone}%
\theoremstyle{thmstyletwo}%

\theoremstyle{thmstylethree}%

\begin{document}

\title[Article Title]{Diffusion Models Bridge Deep Learning and Physics in ENSO Forecasting}


\author[1]{\fnm{Weifeng} \sur{Xu}}
\equalcont{These authors contributed equally to this work.}

\author[1]{\fnm{Xiang} \sur{Zhu}}
\equalcont{These authors contributed equally to this work.}

\author[1]{\fnm{Xiaoyong} \sur{Li}}
\equalcont{These authors contributed equally to this work.}
\author[1]{\fnm{Qiang} \sur{Yao}}
\equalcont{These authors contributed equally to this work.}

\author*[1]{\fnm{Xiaoli} \sur{Ren}}\email{renxiaoli18@nudt.edu.cn}
\author*[1]{\fnm{Kefeng} \sur{Deng}}\email{dengkefeng@nudt.edu.cn}

\author[1, 2]{\fnm{Song} \sur{Wu}}
\author[1]{\fnm{Chengcheng} \sur{Shao}}

\author[3]{\fnm{Xiaolong} \sur{Xu}}
\author[1]{\fnm{Juan} \sur{Zhao}}
\author[1]{\fnm{Chengwu} \sur{Zhao}}
\author[1]{\fnm{Jianping} \sur{Cao}}
\author[1]{\fnm{Jingnan} \sur{Wang}}

\author[1]{\fnm{Wuxin} \sur{Wang}}
\author[1]{\fnm{Qixiu} \sur{Li}}
\author[4]{\fnm{Xiaori} \sur{Gao}}
\author[5]{\fnm{Xinrong} \sur{Wu}}

\author[1]{\fnm{Huizan} \sur{Wang}}
\author*[1]{\fnm{Xiaoqun} \sur{Cao}}\email{caoxiaoqun@nudt.edu.cn}
\author[1]{\fnm{Weiming} \sur{Zhang}}
\author[1]{\fnm{Junqiang} \sur{Song}}
\author*[1]{\fnm{Kaijun} \sur{Ren}}\email{renkaijun@nudt.edu.cn}

\affil*[1]{\orgdiv{College of Meteorology and Oceanography}, \orgname{National University of Defense Technology}, \city{Changsha}, \country{China}}

\affil[2]{\orgdiv{School of Computer Science}, \orgname{Hunan First Normal University}, \city{Changsha}, \country{China}}

\affil[3]{\orgdiv{School of Software}, \orgname{Nanjing University of Information Science and Technology}, \city{Nanjing}, \country{China}}

\affil[4]{\orgdiv{Navigation College}, \orgname{Dalian Maritime University}, \city{Dalian}, \country{China}}

\affil[5]{\orgdiv{National Marine Data and Information Service}, \city{Tianjin}, \country{China}}


\abstract{Accurate long-range forecasting of the El \Nino-Southern Oscillation (ENSO) is vital for global climate prediction and disaster risk management. Yet, limited understanding of ENSO's physical mechanisms constrains both numerical and deep learning approaches, which often struggle to balance predictive accuracy with physical interpretability. Here, we introduce a data driven model for ENSO prediction based on conditional diffusion model. By constructing a probabilistic mapping from historical to future states using higher-order Markov chain, our model explicitly quantifies intrinsic uncertainty. The approach achieves extending lead times of state-of-the-art methods, resolving early development signals of the spring predictability barrier, and faithfully reproducing the spatiotemporal evolution of historical extreme events. The most striking implication is that our analysis reveals that the reverse diffusion process inherently encodes the classical recharge-discharge mechanism, with its operational dynamics exhibiting remarkable consistency with the governing principles of the van der Pol oscillator equation. These findings establish diffusion models as a new paradigm for ENSO forecasting, offering not only superior probabilistic skill but also a physically grounded theoretical framework that bridges data-driven prediction with deterministic dynamical systems, thereby advancing the study of complex geophysical processes.}




\maketitle
\section{Introduction}\label{sec1}
The El \Nino-Southern Oscillation (ENSO) is the dominant mode of interannual climate variability, arising from coupled ocean-atmosphere dynamics in the equatorial Pacific~\cite{cai_pantropical_2019}. It manifests as quasi-periodic fluctuations in sea surface temperature (SST) between neutral, El \Nino (warm phase), and La Ni\~na (cold phase) conditions, exerting profound global impacts on seasonal anomalies and extreme events~\cite{2023Enhanced,hobeichi_how_2024}.
Improving ENSO forecasts is therefore central to advancing early-warning systems, reducing climate-related disaster risks, and strengthening socio-economic resilience~\cite{timmermann_ninosouthern_2018,mcphaden_enso_2006}.

Traditional forecasting approaches rely largely on numerical models that integrate the governing equations of geophysical fluid dynamics through coupled atmosphere-ocean frameworks. These include intermediate coupled models~\cite{zhang_iocas_2016,song_intermediate_2018} 
 and hybrid dynamical systems~\cite{hu_hybrid_2019},
which have deepened understanding of ENSO. However, their predictive performance is limited by sensitivity to parameterizations~\cite{dommenget_analysis_2013} and initial conditions~\cite{zhang_review_2020}. Furthermore, the absence of a unified theoretical framework for ENSO evolution~\cite{liao_enso_2021}, 
compounded by its multiscale nonlinear interactions and decadal modulation, fundamentally constrains model representational capacity. Therefore, it manifests as persistent difficulties in capturing the full spectrum of ENSO event trajectories.


Recent advances in machine learning have introduced promising data-driven alternatives. Early deep learning models, such as convolutional neural networks~\cite{ham_deep_2019} 
and long short-term memory architectures~\cite{yu_responses_2023,chen_toward_2025} 
focused on predicting single ENSO indices. More sophisticated frameworks, including Transformers~\cite{wang_deep_2025,zhou_self-attentionbased_2023,chen_toward_2025} 
 and graph neural networks~\cite{liang_adaptive_2024}, have since been developed to account for ENSO's spatiotemporal complexity. To improve interpretability, researchers have also embedded physical knowledge into learning systems through feature engineering~\cite{chen_toward_2025,wang_role_2024}, physics-informed losses~\cite{wu_explainable_2025}, 
 and physically constrained architectures~\cite{mu_incorporating_2024}. Nonetheless, current approaches still struggle to fully elucidate critical physical processes, such as the onset of the spring predictability barrier (SPB) and its governing mechanisms~\cite{chen_toward_2025}.


Both numerical and machine learning models share a fundamental limitation. They establish deterministic mappings from historical to future states. Numerical models approximate physical laws but are constrained by simplifications and parameterizations, 
while machine learning approaches extract statistical correlations without guarantees of physical consistency~\cite{runge_inferring_2019}. 
The nonlinear dynamics of the climate system exhibit intrinsic chaos~\cite{mcphaden_enso_2006}, particularly in the ENSO cycle where stochastic forcings interact with deterministic feedbacks. The models even harder to capture the initiation criteria and phase-transition pathways of ENSO events. Consequently, purely deterministic mappings inadequately represent ENSO-related uncertainties, and skillful forecasts beyond two-year lead times remain elusive~\cite{mcphaden_enso_2006}.

\begin{figure}[htbp]    
	\centering            
	\includegraphics[width=1.0\linewidth]{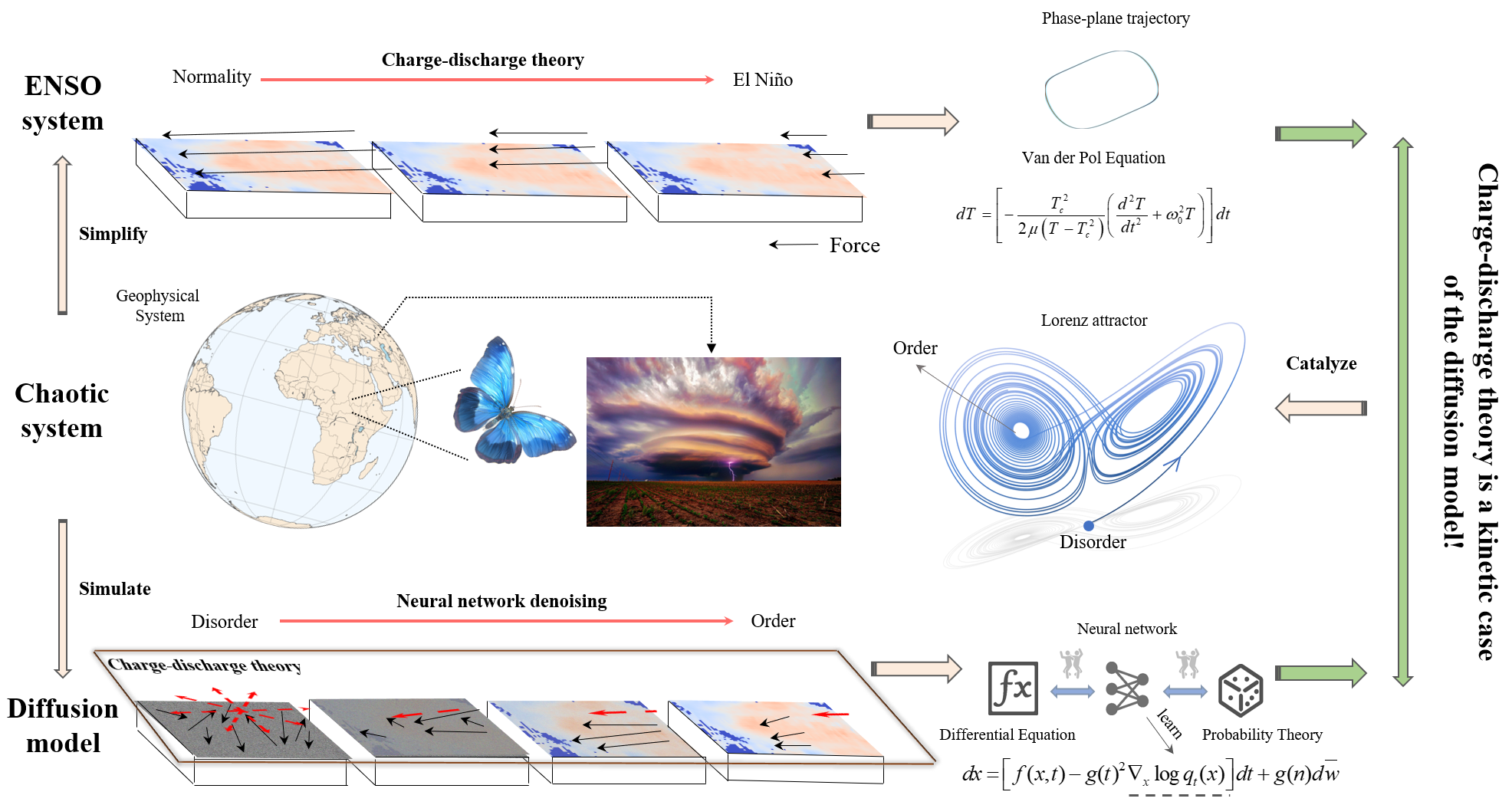}
	\caption{
		\textbf{Triadic Linkage of Recharge-Discharge Physics, Chaotic Dynamics, and Diffusion Reverse-Time SDE for Interpretable ENSO Prediction.}}
	\label{fig:12}   
\end{figure}

Diffusion models, which combine stochastic differential equations with neural networks~\cite{karras_elucidating_2024,song_score-based_2020}, offer a new probabilistic framework that explicitly quantifies uncertainty~\cite{ji_toward_2025,li_generative_2024,price_probabilistic_2025}  
while retaining physical interpretability. Here we  modify the diffusion model for ENSO prediction performance, and demonstrate the model inherently encodes the recharge-discharge mechanism in a form mathematically equivalent to the van der Pol oscillator. This dual achievement establishes diffusion models as a physically consistent and transferable framework for ENSO prediction, bridging data-driven learning with deterministic dynamical systems, shown in Fig.\ref{fig:12}.

\section{Extended Long-Range Forecasting Skill with Diffusion Models}\label{sec2}
To systematically evaluate the predictive performance of the diffusion model, we focus on the \Nino3.4 region ($5^{\circ} N-5^{\circ} S, 170^{\circ}-120^{\circ} W $), the standard benchmark for ENSO diagnostics. Forecast skill is assessed using two widely adopted metrics: the anomaly correlation coefficient of the \Nino3.4 index and the mean absolute error (MAE) of sea surface temperature anomalies (SSTA)~\cite{ham_deep_2019,chen_toward_2025,zhou_self-attentionbased_2023,wang_role_2024,chen_combined_2025}. The dataset spans 1980-2021, providing a comprehensive basis for validation.

\begin{figure}[htbp]    
	\centering            
	\includegraphics[width=1.0\linewidth]{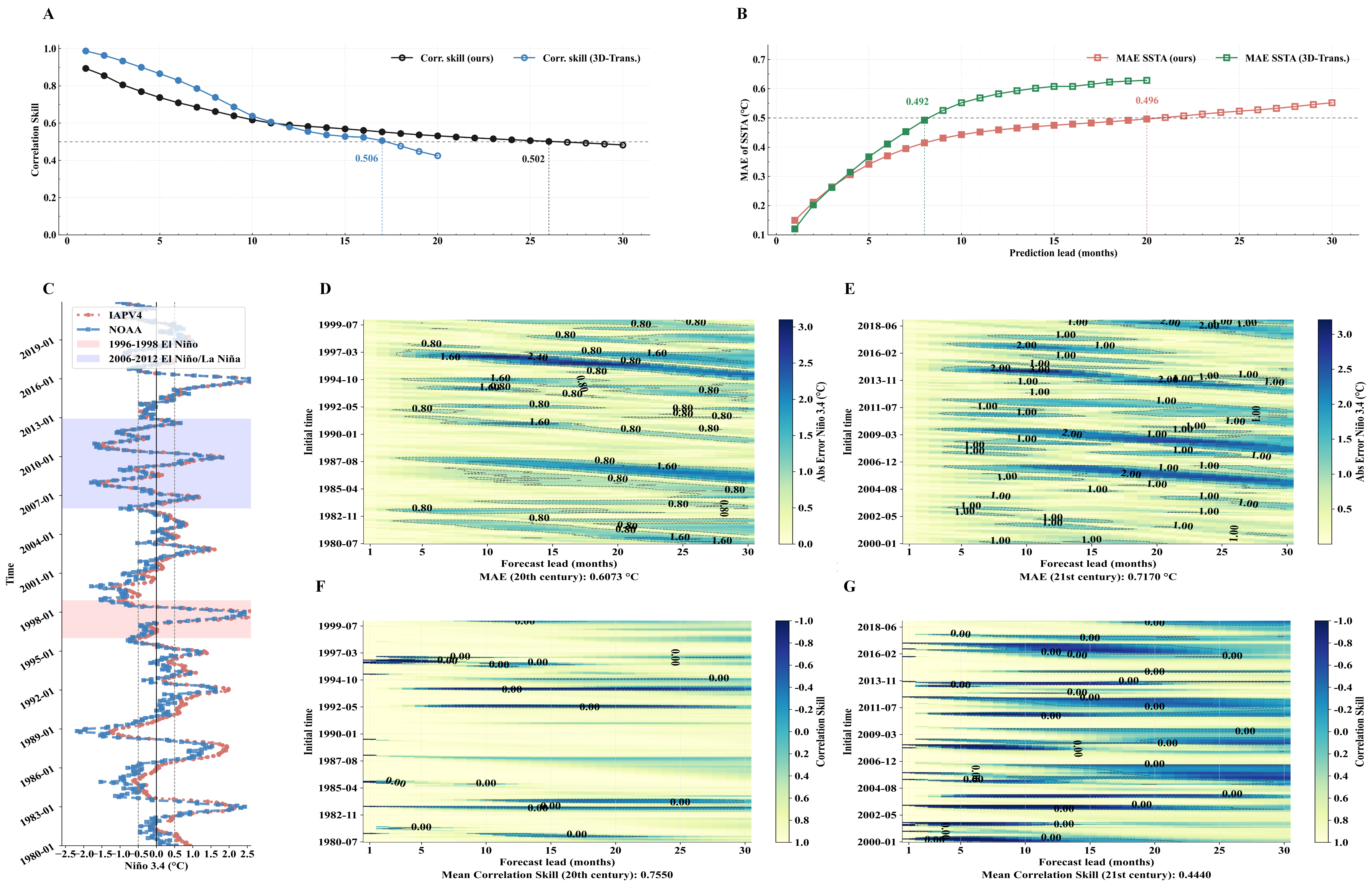}
	\caption{\textbf{Performance of the diffusion model in 30-month-ahead ENSO prediction from 1980 to 2021.}
		\textbf{A-B}: Average evolution of evaluation metrics across lead times (1-30 months) during 1980-2021. Dots mark skill scores; filled dots indicate values exceeding 0.5. Squares denote the mean absolute error (MAE) of SSTA; filled squares correspond to MAE below 0.5$^{\circ}$C.
		\textbf{C}: Comparison of \Nino3.4 index time series between the IAPV4 dataset and NOAA observations from 1980 to 2021 to test whether IAPV4 reproduces the observed ENSO amplitude phase and decadal swings from 1980 to 2021. The two curves show strong agreement with an overall correlation coefficient of 0.9617.
		\textbf{D} (20th century) and \textbf{E} (21st century): Heat-maps of absolute error for every \Nino{3.4} sample. The x-axis indicates lead time, and the y-axis represents prediction start time.
		\textbf{F} (20th century) and \textbf{G} (21st century): Heatmap of correlation coefficients for all samples. The x-axis indicates lead time, and the y-axis represents prediction start time. } 
	\label{fig:1}   
\end{figure}

As shown in Fig.\ref{fig:1}A, the diffusion model sustains correlation skill above 0.5 for up to 26 lead months well beyond the useful range of most operational dynamical and machine-learning forecasts~\cite{chen_combined_2025,ham_deep_2019,zhou_self-attentionbased_2023,mu_incorporating_2024,zhao_explainable_2024}.  
Over the same horizon, the monthly MAE of SSTA increases only gradually, from 0.15 $^{\circ}$C to 0.55 $^{\circ}$C, with limited variability. Unlike conventional studies that report aggregated scores~\cite{ham_deep_2019,wang_role_2024,jiang_climate_2024}, 
we provide detailed lead-dependent time series of both correlation and error, allowing a more granular assessment of forecast evolution and reducing distortion from high-variance regimes or outliers. Aggregated across the entire horizon, the model achieves a 30-month mean MAE of 0.66 $^{\circ}$C for the \Nino3.4 index. For comparison, the 6-9 month mean El \Nino bias exceeds 0.7$^{\circ}$C for both DYN and STAT forecasts from the the International Research Institute for Climate and Society multi-model ensemble mean~\cite{ehsan_real-time_2024}. The diffusion model thus not only extends reliable forecast guidance by over two years but also maintains error levels competitive with leading dynamical systems at their optimal leads.

Fig.\ref{fig:1}D-G reveal a decline in forecast skill after 2000, coinciding with enhanced interdecadal ENSO variability. This variability intensifies interactions between interannual and low-frequency modes, producing more heterogeneous SST patterns that challenge generalization, particularly for models trained on relatively stationary regimes.  Studies~\cite{chen_improved_1995,tang_progress_2018} 
confirm that such non-stationarity degrades predictive fidelity, while numerical models struggle to represent multi-scale processes such as MJO-ENSO interactions and trans-basin teleconnections~\cite{zhang_recent_2022}. The observed post-2000 performance decline thus reflects both a shifting climate baseline and the limited physical realism of available training datasets.

Periods of degraded performance often align with challenging events, such as the 1997-1998 super El \Nino and the rapid El \Nino-La Ni\~na transitions of 2006-2012. These intervals feature strong nonlinearities and rapid phase shifts. Nevertheless, the diffusion model shows marked improvement when initialized with seasonally critical windows. For instance, using SST from March-August 1997 as input reduces \Nino3.4 prediction error relative to earlier initializations, enabling more accurate capture of the extreme 1998 El \Nino. This illustrates that incorporating seasonally sensitive signals enhances model skill in tracking ENSO dynamics.

\section{Observation-Based Training Improves ENSO Forecasting in the 2lst Century}\label{sec3}
The analysis in Section~\ref{sec2} shows that diffusion models pre-trained on the CMIP6 multi-model ensemble exhibit limited skill in predicting 21st-century ENSO events. This limitation may reflect either structural constraints of the diffusion framework or systematic biases in CMIP6-based ENSO simulations~\cite{zhang_recent_2022,fu_simulated_2021}.  
To disentangle these factors and improve forecast reliability, we adopt a de-modeling strategy: rather than relying on potentially biased model outputs, we train the diffusion model directly on high-resolution observational reanalysis data.
Specifically, the model was trained on the IAP v4 global SST reanalysis from 1940 to 1999, a dataset that integrates multiple observational sources and accurately captures historical SST variability and ENSO characteristics~\cite{cheng_iapv4_2024,wang_strong_2025}. 
Independent validation was performed using IAP v4 data from 2000 to 2023.

\begin{figure}[htbp]    
	\centering            
	\includegraphics[width=1\linewidth]{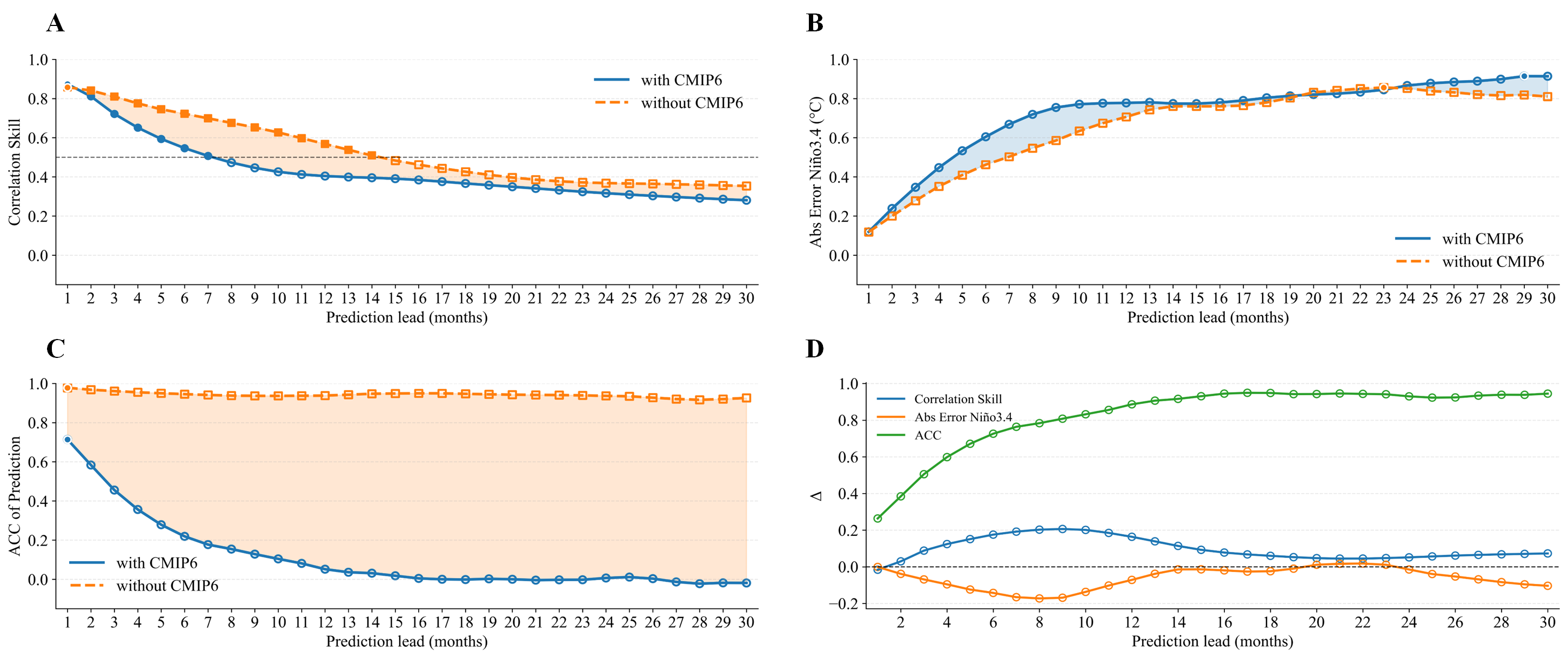}
	\caption{\textbf{Prediction Performance Comparison Between Pre-trained and Non-Pre-trained Models (2000-2023)}
	\textbf{A-C}: Average evolution of evaluation metrics across lead times (1-30 months) during 2000-2023. \textbf{A} filled dots mark correlation coefficients $>$ 0.5. \textbf{B} filled dots indicate the maximum absolute \Nino3.4 index error in each forecast.
	\textbf{C} filled dots denote the maximum spatial correlation coefficient achieved in each prediction.
	\textbf{D}: Difference (without-CMIP6 minus with-CMIP6) for each metric, highlighting the impact of CMIP6 training data.} 
	\label{fig:2}   
\end{figure}

Fig.\ref{fig:2} compares hindcast performance with and without CMIP6 pre-training. The observation-only model extends the effective \Nino3.4 correlation horizon to 14 months double that of the CMIP6-pretrained counterpart, which achieves only 7 months. Spatial fidelity shows a similar pattern: the CMIP6-free model preserves coherent ENSO-related structures across the forecast horizon, whereas the pre-trained model displays degraded spatial correlations, reflecting reduced adaptability to evolving ENSO modes. Although overall error magnitudes are comparable, the observation-driven model consistently produces smaller \Nino3.4 absolute errors, confirming that reliance on observational baselines yields sharper and more reliable forecasts.

These results underscore the decisive role of training data in shaping diffusion model skill. The IAP v4 dataset~\cite{cheng_iapv4_2024} provides a faithful representation of SST variability and ENSO dynamics, enabling the diffusion model to internalize physically consistent processes relevant to contemporary climate. By contrast, CMIP6 pre-training introduces systematic biases, forcing the network to unlearn model-imposed artifacts before adapting to observational reality. This process reduces both efficiency and accuracy. Eliminating CMIP6 pre-training allows the model to focus exclusively on real-world ENSO dynamics, confirming that performance degradation arises primarily from dataset biases rather than architectural limitations.

\section{Probabilistic Modeling Reveals Early Signals of the Spring Predictability Barrier}\label{subsec2}
\begin{figure}[htbp]    
	\centering            
	\includegraphics[width=1.0\linewidth]{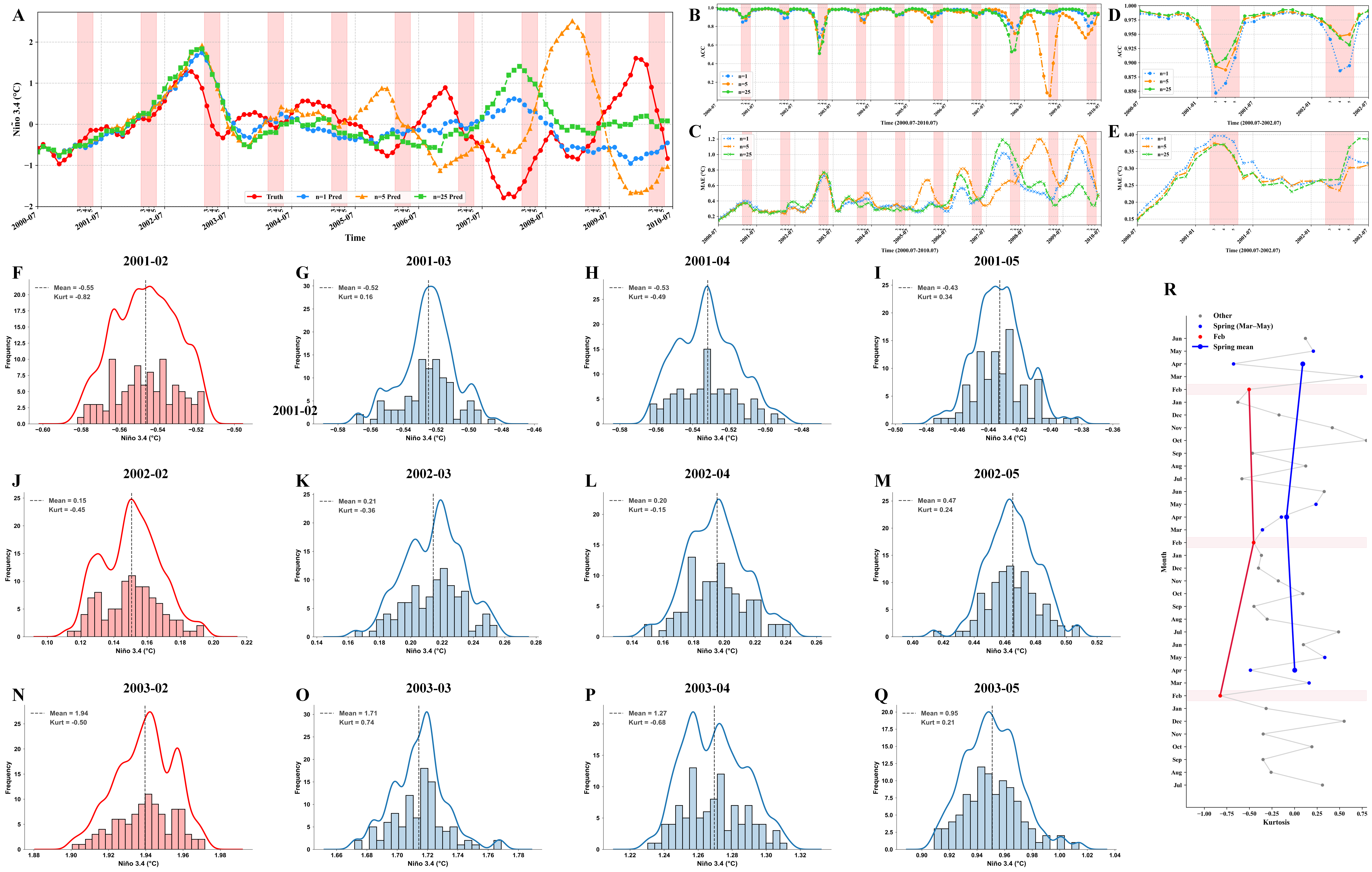}
	\caption{\textbf{Model Performance under the Spring Predictability Barrier for ENSO. Red shading marks spring.} 
		\textbf{A-E}: Hindcast skill from January-June 2000 initial conditions for the period July 2000-July 2010, $n$ denotes the ensemble size.
		\textbf{F-Q}: Distributions of the monthly \Nino3.4 index produced with 100-member ensembles (n = 100). Histograms and probability-density curves are obtained by kernel-density estimation (bandwidth = 0.2).
		\textbf{R}: Kurtosis for July 2000-June 2003. The blue line shows the spring-average kurtosis, red symbols indicate values for the February immediately preceding the Spring Predictability Barrier.}
	\label{fig:3}   
\end{figure}

The Spring Predictability Barrier (SPB) is a well-known phenomenon characterized by an abrupt and systematic decline in ENSO forecast skill during boreal spring~\cite{chen_central-pacific_2024,yu_potential_2024,landman_verification_2025}, posing a major challenge for operational ENSO prediction, as it substantially reduces the reliability of forecasts spanning this period ~\cite{gupta_prediction_2020,meng_complexity-based_2020}. To investigate the ability of diffusion models to address this challenge, we employ the IAP v4-trained model to quantify how ensemble size modulates predictive uncertainty across multiple SPB events. 
Specifically, we initialize the model with January-June 2000 SST fields and generate 100-member ensembles for forecasts extending through June 2003, thereby enabling a detailed assessment of uncertainty dynamics around the SPB.
The histograms and probability density functions of the \Nino3.4 index (Fig.\ref{fig:3}F-Q) capture the evolving kurtosis across multiple SPBs. Particular attention is paid to the February forecast outputs before the barrier to isolate early warning signals of the impending predictability collapse.

The model demonstrates robust ENSO forecasting skill from July 2000 to November 2002, accurately reproducing \Nino3.4 index evolution across different ensemble configurations while maintaining high spatial fidelity and low mean absolute error (MAE$<$1.2$^{\circ}$C). However, predictive performance deteriorates markedly from August 2002 to 2003, especially in amplitude estimation, with pronounced phase lags emerging in the May-July 2003 trough prediction. The spring 2003 period reveals critical limitations when the SPB interacts with strong El \Nino\ conditions~\cite{mcphaden_evolution_2004}, manifesting as collapsed spatial correlations and substantially elevated errors. Increasing the ensemble size beyond five provides limited gains, making mid-sized ensembles optimal for balancing the accuracy and efficiency. These results highlight both the model's strengths and its limitations in capturing ENSO variability across SPB windows.


Importantly, the model successfully reproduces the canonical life cycle of the SPB and quantifies the rapid increase in predictive uncertainty accompanying its onset. Across multiple years, ensemble distributions display pronounced dispersion in each February, reflected in a sharp decline in kurtosis relative to other months. This flattening signals a critical transition from a tightly clustered predictive density in the preceding months to a markedly broader distribution just before the barrier sets in. The transition itself constitutes an unambiguous early-warning signature generated internally by the model. Once the core barrier window arrives, the ensemble spread relative to the February kurtosis rebounds, indicating the model can still retain relatively usable predictive information within the barrier period, consistent with the results shown in Fig.\ref{fig:3}B-E. The model's keen sensitivity to uncertainty changes at the onset of SPB is its standout feature, representing a key advance in tackling the SPB challenge.

Overall, the primary merit of the diffusion model lies in its capacity to generate large ensembles that objectively quantify evolving forecast uncertainty. Specifically, it detects the pronounced uncertainty surge before the onset (February). This capability furnishes a quantitative framework for mechanistic dissection of the SPB and, more critically, enables operational centers to identify barrier periods weeks to months in advance while objectively quantifying forecast risk, thereby enhancing SPB early-warning capacity and mitigating its detrimental impact on ENSO forecast lead time and reliability.

\section{Reproducing the Spatiotemporal Dynamics of Extreme ENSO Events}\label{subsubsec2}

To quantify the advantages of the diffusion model in ENSO prediction, we adopted the version trained directly on the IAPv4 dataset and selected two extreme cases for in-depth analysis. These cases provide a vivid, side-by-side demonstration of the model's skill and fidelity when confronted with the full complexity of the climate system.

\textbf{The super El \Nino\ during the 2015.}
The 2015 super El \Nino\ ranks among the three strongest events recorded since 1951 and displays a clear central-Pacific signature. It exerted far-reaching impacts on global climate, pushing 2015 to become the warmest year on record, about 0.76$^{\circ}$C above the 1961-1990 average~\cite{wmo_wmo_2016}, while nations around the Pacific endured a series of extreme weather events. Multiple mechanisms contributed to this event, including Bjerknes positive feedback, the delayed oscillator, and the recharge-discharge paradigm, each playing a pivotal role in its formation and evolution~\cite{chen_strong_2015,xie_unusual_2020}.

\begin{figure}[htbp]    
	\centering            
	\includegraphics[width=1.0\linewidth]{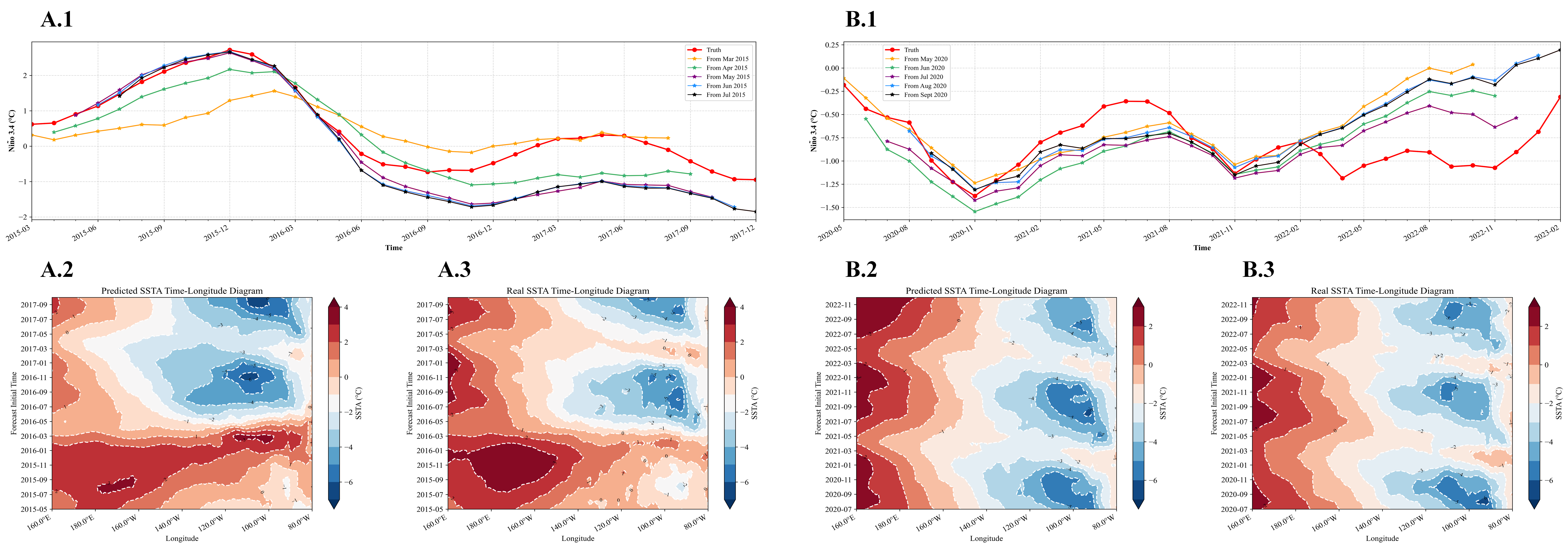}
	\caption{\textbf{Forecast results for the extreme cases.} \textbf{A} for 2015/16 super El \Nino, \textbf{B} for 2020-22 triple La Ni\~na; A.1\&B.1 show the predicted \Nino3.4 index, A.2-3\&B.2-3 display the equatorial spatio-temporal evolution of SST anomalies.}
	\label{fig:5}   
\end{figure}

Fig.\ref{fig:5}A.1 shows that the predicted \Nino 3.4 index closely follows observations in overall trend. The model successfully captures key evolutionary features, including the early 2015 warming phase, peak intensity from late 2015 to early 2016, and subsequent decline. Notably, the predicted curve aligns strongly with observations during the mature phase of 2015-2016. However, forecast accuracy decreases during the decay phase (mid-to-late 2016). The model also shows sensitivity to initialization data: variations in the first six months of input lead to divergent forecasts, mainly due to the climate system's phase at initialization and background conditions. These seasonal differences in oceanic and atmospheric states influence subsequent trajectories of climate evolution~\cite{kim_seasonal_2021,zhen_inter-seasonal_2025}. 
In spatial characteristics (Fig.\ref{fig:5}A.2-A.3), the model reliably reproduces the large-scale SSTA layout, yet finesse is lacking in pinpointing the exact edges, amplitude, and position of anomalies. Its skilful capture of the overall warm-pool geometry notwithstanding, the 2015-16 event exposes timing flaws: the simulated eastward push and intensification trail observations, and the June-September surge in both expansion speed and peak warming is underestimated.

\textbf{The triple La Ni\~na event from 2020 to 2022.}
The \enquote{triple-dip} La Ni\~na emerged without a strong precursor El \Nino, anchored its cold tongue in the far southeast Pacific, and coupled only weakly to the northern subtropics, rendering duration, intensity and pattern unusually hard to predict~\cite{chen_understanding_2025,geng_increased_2023}. 



The model's 2020-2022 \Nino3.4 hindcast tracks the observed triple-dip La Ni\~na within \(\pm0.3\,^{\circ}\mathrm{C}\), shows no systematic drift and correctly anticipates the re-intensification of the first two cold episodes (Fig.\ref{fig:5}B.1). Large-scale patterns are also skilful: the equatorial cold-tongue contraction and the western-Pacific warm-pool anomalies evolve in close agreement with observations (Fig.\ref{fig:5}B.2-B.3), implying that the training encoded the governing air-sea feedbacks. Deficiencies remain in the amplitude of sub-seasonal variability (early 2021 to early 2022) and in the exact location and intensity of boundary transition zones and secondary anomaly centres.


Overall, the diffusion model shows the ability to reproduce the \Nino3.4's long-term trajectory and the main spatial SSTA patterns, highlighting its potential for handling complex ocean-atmosphere interactions. However, its performance in forecast accuracy, fine-scale spatial detail, and temporal alignment still needs improvement and requires further refinement.

\section{Reverse Diffusion Encodes Recharge-Discharge Dynamics of ENSO}\label{7}
\begin{figure}[htbp]    
	\centering            
	\includegraphics[width=1.0\linewidth]{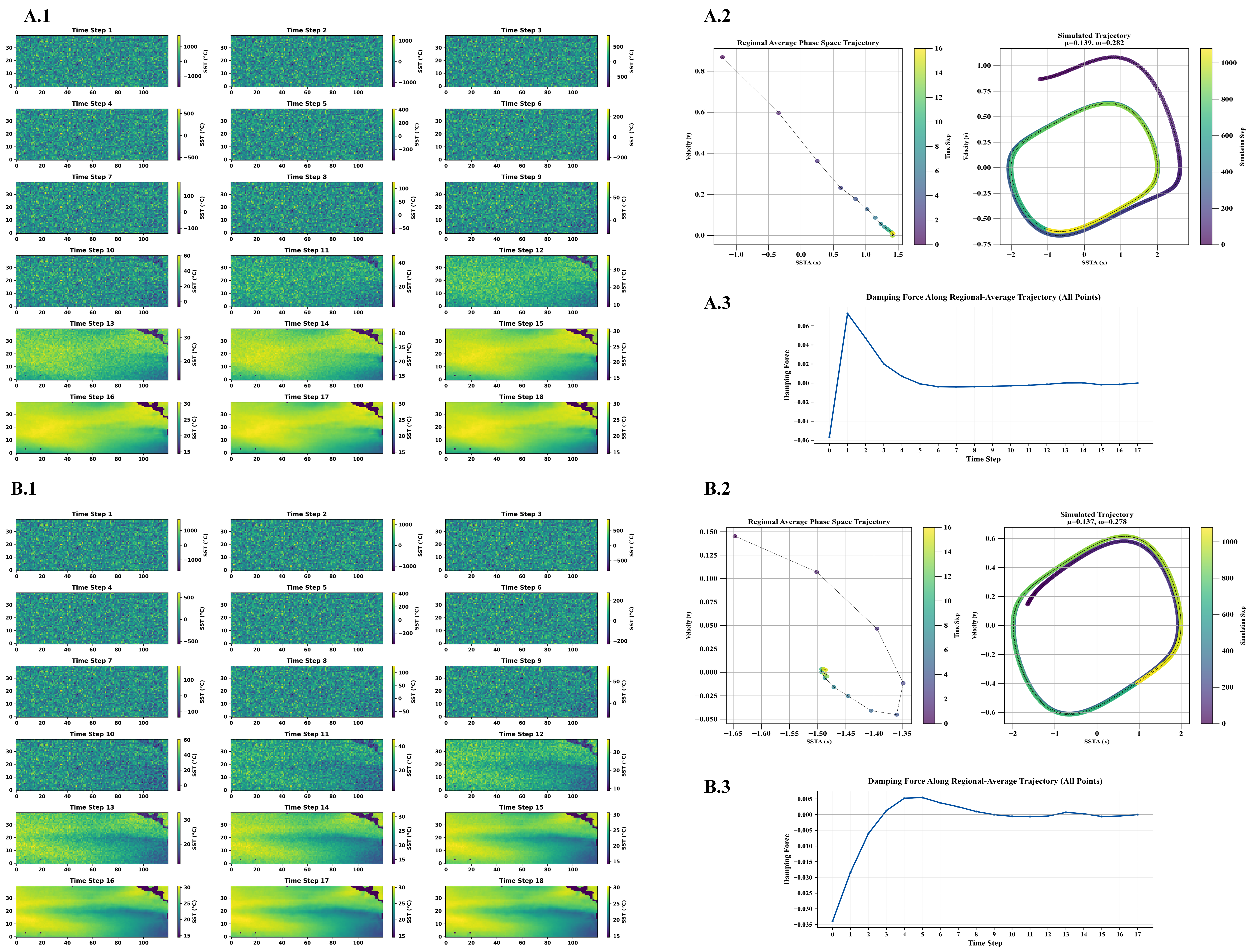}
	\caption{\textbf{Analysis of Van der Pol model fitting and dynamic evolution for \Nino3.4 region's mean sea surface temperature anomalies.} The figure contains two case series: Panel A (July 2015 event) and Panel B (November 2020 event).\textbf{A.1 \& B.1}: Spatiotemporal evolution of the 18-step reverse diffusion process. \textbf{A.2 \& B.2}: Phase-space trajectories from model simulations, with color gradients indicating simulation steps. The left shows observed phase-space trajectories from real-world data, while the right displays 5-year projections simulated using the fitted Van der Pol equation. \textbf{A.3 \& B.3}: Temporal evolution of nonlinear damping forces across simulation steps.} 
	\label{fig:8}   
\end{figure}

To assess the consistency between the reverse process of the diffusion model and the physical dynamics of ENSO-particularly the applicability of the Van der Pol equation within the recharge-discharge oscillator framework~\cite{wu_explainable_2025,wang_dual-core_2025}, 
we analyze prediction results from July 2015 and November 2020. These cases represent critical ENSO phases: July 2015 corresponds to the rapid development of a strong El \Nino, while November 2020 marks the secondary peak of a double-peak La Ni\~na event. Both are characterized by nonlinear phase transitions and high predictive uncertainty, making them ideal testbeds. We visualized the 18-step reverse inference process (Fig.\ref{fig:8}). For the \Nino3.4 region, we assume each grid point follows the Van der Pol equation over these steps and estimate parameters accordingly. The coefficient of determination ($R^2$) measures the fraction of acceleration variance explained by the van-der-Pol model, thereby directly assessing the alignment between the model's reverse dynamics and ENSO's governing physics (Fig.\ref{fig:8}).

\begin{table}[h]
	\centering
	\caption{Van der Pol model fit statistics for \Nino3.4 region (July 2015 \& November 2020)
		Mean nonlinear damping $\mu$, mean oscillation frequency $\omega$ and mean goodness-of-fit $R^2$ are reported as mean $\pm$ standard deviation.}\label{tab3=}%
	\begin{tabular}{lll}
		
		\toprule
		 & 2015.7  & 2020.11 \\
		\midrule
		Mean nonlinear damping $\mu$    & 0.1388 $\pm$ 0.0261   & 0.1365 $\pm$ 0.0512   \\
		Mean oscillation frequency $\omega$    & 0.2824 $\pm$ 0.0775   & 0.2777 $\pm$ 0.0877  
		\\
		Mean goodness-of-fit $R^2$    & 0.9916 $\pm$ 0.0557   & 0.9948 $\pm$ 0.0243 
		\\
		Mean trajectory discrepancy $\Delta$ &0.3697&0.3762\\
		Correlation coefficient ($\mu$ vs $\omega$)&0.9236&0.5879
		\\
		Correlation coefficient ($\mu$ vs $R^2$)&0.5779&0.8293\\
		
		\botrule
	\end{tabular}
\end{table}

Fig.\ref{fig:8} illustrates the inference phase of the diffusion model. Based on the past six months of observations, it progressively denoises random noise into a forecast of the month ahead. A trained neural network guides each step, turning a blur of Gaussian noise into the final, structured signal.
Table~\ref{tab3=} confirms that this denoising process is not merely statistical mimicry. The reverse-time equations of the diffusion model align almost perfectly with the van der Pol oscillator, so closely that they effectively embed the ENSO oscillator itself. Thus, the model's reverse pass is physically grounded.
A residual trajectory mismatch remains. It could be due to training error, or it may simply mean the diffusion model captures ENSO more faithfully than the classical oscillator.
Finally, the recovered parameters-non-linear damping $\mu$ and natural frequency $\omega$-reproduce the observed self-limiting amplitude and a dominant $~$22-month cycle, underscoring the model's capacity to mirror ENSO's essential physics.

Notably, $\mu$ and $\omega$ exhibit a strong positive correlation with clear physical meaning. Larger $\mu$ reflects intensified nonlinear feedbacks in the ENSO system, which accelerate oceanic adjustment through rapid eastward propagation or westward retreat of thermocline anomalies. Faster adjustment shortens the oscillation period, thereby increasing $\omega$. This finding reinforces the self-regulating nature of ENSO and underscores the ability of the diffusion-Van der Pol framework to represent complex nonlinear dynamics. Moreover, $\mu$ positively covaries with $R^2$, indicating that stronger nonlinear feedbacks improve the oscillator's ability to reproduce \Nino3.4 variability. This result is consistent with the dominance of nonlinear advection and strong air-sea coupling in shaping SST anomalies in the \Nino3.4 region.


Fig.\ref{fig:8} presents three paired panels that jointly illustrate the diffusion model's ability to reproduce the temporal unfolding of ENSO. The trajectories generated by the reversed SDE match the observed phase-space evolution, sharing not only the same directional trends but also the underlying dynamic origin. This congruence substantiates the physical legitimacy of the reversed SDE for capturing ENSO's core oscillation between cold and warm states.
The entire evolution is orchestrated by a smooth transition of the damping force from negative to positive values. This transition maps one-to-one onto the canonical ENSO sequence-cold decay, warm onset, warm peak, warm decay, and cold re-emergence-thereby embodying the slow accumulation and rapid release that define relaxation oscillations. Moreover, the gradual convergence of the damping coefficient over successive time steps exposes a deeper mathematical property: the system settles into a limit cycle sustained by an exact balance between energy injection (negative damping) and dissipation (positive damping). Consequently, the reversed SDE not only replicates individual ENSO events but also encodes the long-term, self-sustaining cyclicity of the system. The three panels in Fig.\ref{fig:8}, therefore, provide compelling visual evidence that the diffusion-based reversed SDE faithfully represents the essential dynamics of ENSO.

In summary, the reverse-time SDE formulation of the diffusion model can faithfully approximate the essential dynamical behaviour of the recharge-discharge ENSO oscillator as described by the Van der Pol equation. It suggests that the neural network's training procedure implicitly reconstructs the nonlinear feedback term of the equation, thereby reproducing the complete physical evolution of ENSO. In this context, the diffusion model can be regarded as a physics-guided predictive paradigm, as it not only distils statistical regularities from data but also encodes the underlying dynamical structure, thereby deepening mechanistic insight into ENSO while simultaneously enhancing forecast reliability and physical consistency.

\section{Conclusion}

Diffusion models, formulated as stochastic differential equations (SDEs), employ neural networks to learn the equation structure in a fully adaptive manner, yielding a markedly improved representation of ENSO dynamics.  Even without explicitly embedding physical constraints in the loss function, the reverse-time denoising trajectory reconstructs a complete ``noise $\to$ state'' evolution that is dynamically equivalent to the recharge-discharge cycle of warm-water volume described by the van der Pol oscillator.  This mathematical equivalence provides the model with an intrinsically interpretable account of El \Nino-Southern Oscillation behavior and performs well.  
These contributions, extending beyond ENSO, establishes a reproducible ``physics-AI'' template for application to other chaotic systems.  At its core lies the ability to integrate physical rigor with data-driven efficiency, transforming opaque black-box predictions into an interpretable, dynamics-based generative paradigm.  In Earth system science, this framework can be generalized to multi-sphere couplings, providing an explainable and quantifiable theoretical foundation for anticipating extreme events in a warming climate.

\section{Methods}

\subsection{Task definition}

We reformulate ENSO prediction as a joint-probability learning problem: the network is optimised to recover the full conditional density $P(\textbf{X}_{t+1}|\textbf{\textit{H}})$, where $\textbf{\textit{H}}$ denotes the observed ocean-atmosphere history and $\textbf{X}_{t-1}\in R^d$ the multi-variable future state. Forecasts are delivered as explicit distributions, furnishing native, flow-dependent uncertainty estimates that obviate post-hoc calibration.


To keep the problem tractable without sacrificing physics, we use sea-surface temperature (SST) as a proxy for the state vector $\textbf{\textit{X}}$. SST is the clearest and most directly observed fingerprint of ENSO~\cite{yeh_nino_2009,cai_increased_2022,li_insights_2025}, carrying the essential information while keeping model complexity in check. Consequently, we recast ENSO's evolution as a straightforward statistical task: given the observed record of SST up to the present, estimate the full probability distribution of SST at the next time step.
\begin{equation}
	P\left(T_{t+1}\mid T_t,T_{t-1},T_{t-2},T_{t-3},T_{t-4},T_{t-5}\right).
	\label{definition}  
\end{equation}

Balancing ENSO's physical memory, the point at which forecast skill stops improving, the danger of over-fitting, and the need for computational efficiency, we deliberately truncate history at six months: the prediction for next month is conditioned only on the previous half-year of SST fields.

\subsection{Diffusion model}

\begin{figure}[htbp]    
	\centering            
	\includegraphics[width=1.0\linewidth]{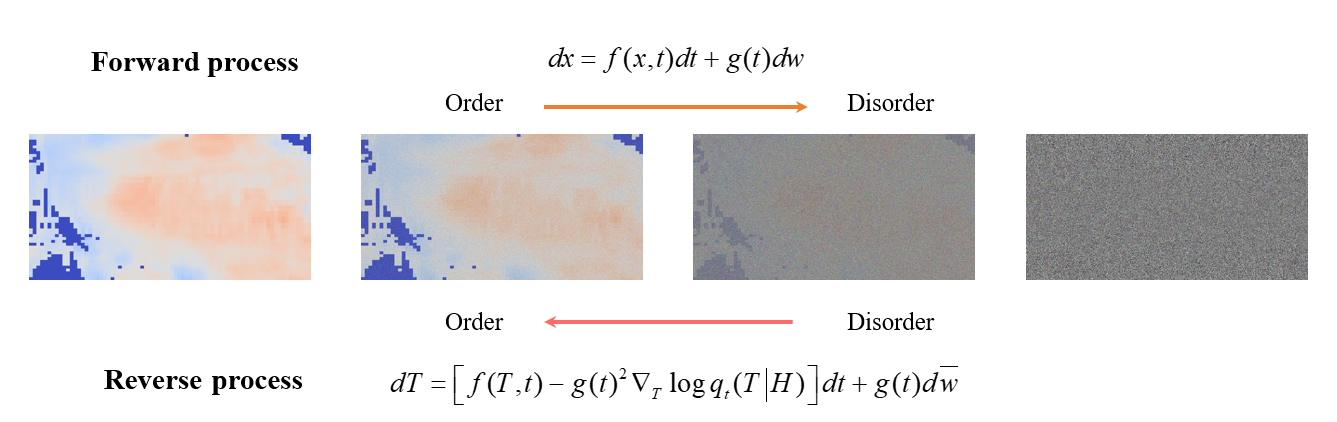}
	\caption{
			\textbf{Schematic illustration of diffusion model.}}
	\label{fig:9}   
\end{figure}

\textbf{Framework.} Diffusion models are a deep Bayesian engine that learns a distribution by gradually reversing an artificial noising process~\cite{yeh_nino_2009,cai_increased_2022,li_insights_2025}. These models have gained widespread application across meteorological and oceanographic domains~\cite{price_probabilistic_2025,li_generative_2024,bassetti_diffesm_2024,ji_toward_2025,zhong_fuxi-extreme_2024,yu_diffcast_2024}.  We split the required conditional density into a sequence of easy-to-fit Gaussian mixtures, each one conditioned on the previous six months of SST.
\begin{equation}
	P\!\left(T_{t+1}\mid H\right)=
	\sum_{k=1}^{K}\pi_k(H)\,
	\mathcal{N}\!\bigl(\mu_k(H),\,\Sigma_k(H)\bigr),
	\quad
	\text{s.t.}\quad
	\sum_{k=1}^{K}\pi_k(H)=1,\; \pi_k(H)\ge 0.
	\label{eq:gaussian_mixture}
\end{equation}
where $\mu_k\left(H\right)$ is the mixture weight of the k-th component given history $\textbf{\textit{H}}$, and  $N\left(\mu_k(H),\sum_k(H)\right)$ is the corresponding Gaussian density, also conditioned on $\textbf{\textit{H}}$.
This factorisation included two stages. First, a prescribed forward noising chain gradually turns the original SST distribution into a simple Gaussian.  Mathematically, this is equivalent to decomposing the data distribution into a sequence of elementary, conditional Gaussian transitions. Second, the network learns the reverse chain: at every step, it predicts the previous, slightly cleaner state from the current, noisier one, and by stringing these learned reversals together, it reconstructs the full SST distribution.   The complete noising-and-denoising loop is sketched in Fig.\ref{fig:9}.

The probability flows of both the forward and reverse processes in the model are continuous-time stochastic differential equations (SDEs)~\cite{song_score-based_2020,song_consistency_2023}, shown in Equations (\ref{eq:sde_forward}) and (\ref{eq:sde_reverse}).
\begin{equation}
	dT = f(T, t) dt + g(t) d\mathbf{w},
	\label{eq:sde_forward}  
\end{equation}

\begin{equation}
	dT = \left[ f(T, t) - g(t)^2 \nabla_{T} \log q_t(T|H) \right] dt + g(t) d\bar{\mathbf{w}},
	\label{eq:sde_reverse}  
\end{equation}

where $f(T, t)$ and $g(t)$ are the drift and diffusion coefficients, $\mathbf{w}$ is a standard Wiener process, $\bar{\mathbf{w}}$ is its time-reversed counterpart, and $dt$ denotes an infinitesimal negative time step.
The score $\nabla_{T} \log q_t(T|H)$ is learned by a deep neural network. 
This continuous-time formulation offers a new lens on ENSO behavior. The forward noising process emulates how initial errors, amplified by air-sea interactions, evolve from structured anomalies into climatic noise. The reverse denoising process uses the network to extract multiscale features, tune the Gaussian-mixture parameters, and encode the dominant physics, guiding the system from chaos back to a physically interpretable, large-scale SST anomaly pattern.

\textbf{Denoiser.} During training, the network learns to recover a future sea-surface temperature field. We first corrupt this target field with a Gaussian noise schedule until it is almost pure noise, keeping every intermediate noisy snapshot as the label the network must predict. Concurrently, the six most recent SST fields are dynamic conditioning, steering the network to estimate the gradient of the conditional log-probability and thereby depicting the distribution $P\left(T_{t+1}\mid T_t,T_{t-1},T_{t-2},T_{t-3},T_{t-4},T_{t-5}\right)$.
The loss is taken verbatim from the EDM framework, given in Equations (\ref{eq:denoiser}) and (\ref{eq:loss}).

\begin{equation}
	\mathcal{D}_\theta(\mathbf{x};\sigma) = c_{\text{skip}}(\sigma) \mathbf{x} + c_{\text{out}}(\sigma) F_\theta\left( c_{\text{in}}(\sigma) \mathbf{x}; c_{\text{noise}}(\sigma) \right),
	\label{eq:denoiser}
\end{equation}

\begin{equation}
	\mathbb{E}_{\sigma,y,n}\!\left[
	\lambda(\sigma)\,c_{\text{out}}^{2}(\sigma)
	\left\lVert
	F_{\theta}\!\left(c_{\text{in}}(\sigma)(y+n);\,c_{\text{noise}}(\sigma)\right)
	-\frac{1}{c_{\text{out}}(\sigma)}\!
	\left(y-c_{\text{skip}}(\sigma)(y+n)\right)
	\vphantom{\frac{1}{c_{\text{out}}(\sigma)}}
	\right\rVert_{2}^{2}\,\right].
	\label{eq:loss}
\end{equation}
where, the input $x = z + n$ is the sum of a clean signal $z$ and Gaussian noise $n~N(0,\sigma^2I)$.
The four scalar functions $c_{\text{in}}(\sigma)$, $c_{\text{skip}}(\sigma)$, $c_{\text{noise}}(\sigma)$, $c_{\text{out}}(\sigma)$,  derived from the noise variance $\sigma^2$, re-scale both the target and the network output so that the loss operates on unit-variance quantities.  stabilises training and improves generalisation (details in~\cite{karras_elucidating_2024,karras}).

\begin{figure}[htbp]    
	\centering            
	\includegraphics[width=1.0\linewidth]{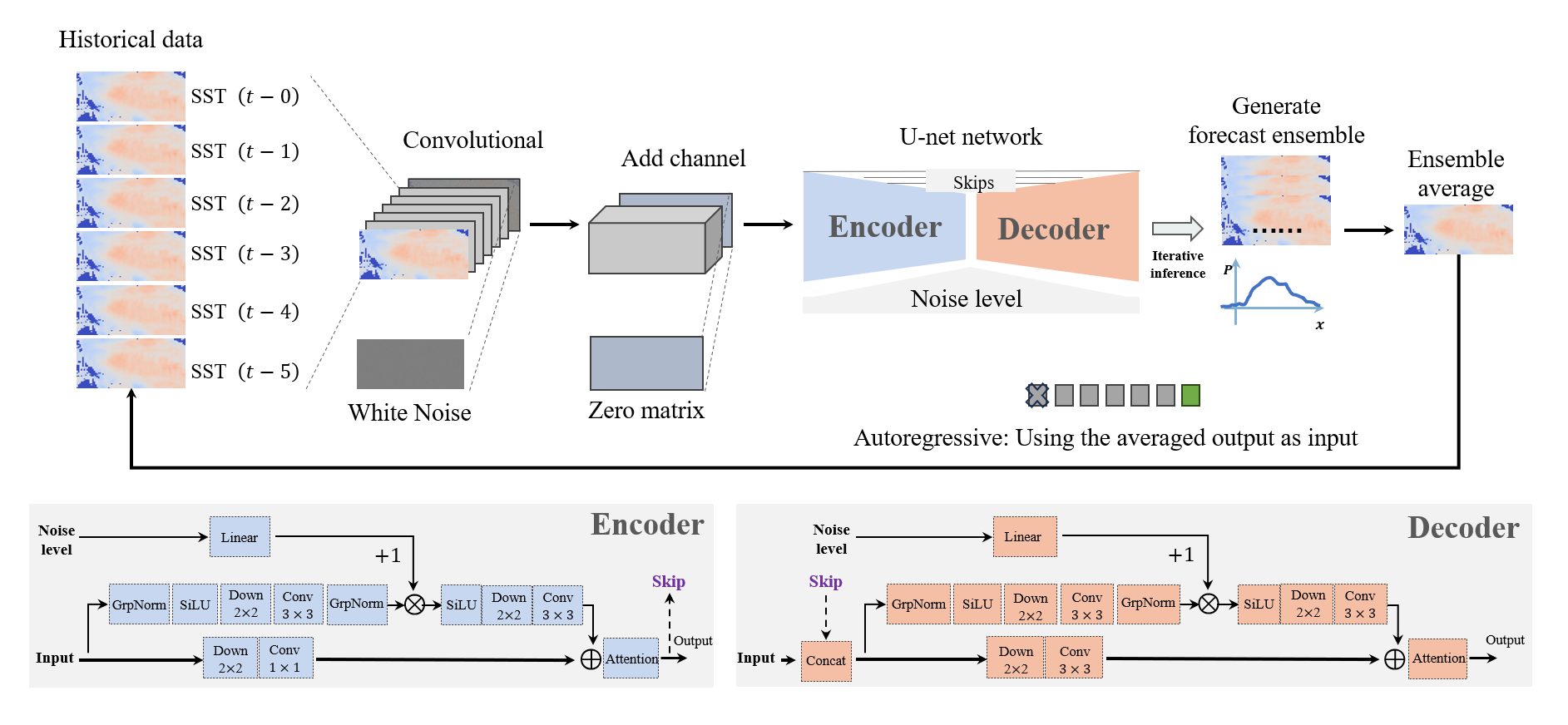}
	\caption{
			\textbf{Framework and network architecture of the diffusion model for ENSO prediction.}}
	\label{fig:10}   
\end{figure}

\textbf{Network architecture.} To honour the 6-step Markov property assumed for ENSO, we modified the standard U-Net (see Fig.\ref{fig:10}).
Classic EDM feeds only the noise level and a class label into the embedding layers, an arrangement ill-suited to a continuous 6-frame SST history. We therefore add a lightweight spatio-temporal conditioning block: the six past SST fields are concatenated with the current noisy target along the channel axis, giving a 7-channel tensor. A $1 \times 1$ convolution then learns a nonlinear channel-mixing map, while an eighth auxiliary channel is left open, giving the model room to capture any additional cross-channel dependencies it deems necessary.

\textbf{Sampling strategy.}
Once training is complete, the model produces an ensemble of high-resolution forecasts by iterating the learned reverse-time steps.
Each member is an independent sample from the predictive distribution, so the pool naturally reveals multiple plausible futures and, without further modeling, gives the probability of any quantile or extreme-event threshold.
In this way, the intrinsic uncertainty of ENSO is displayed explicitly rather than hidden in a single best guess.
From the generated SST fields, we compute the Ni\~no3.4 index by averaging the anomalies over the central-eastern Pacific ($5^\circ\mathrm{S}$--$5^\circ\mathrm{N}$, $170^\circ$--$120^\circ\mathrm{W}$) and thus obtain its full probability distribution. When we need a deterministic benchmark, we take the ensemble mean. To stretch the lead time, we roll the forecast forward: after every month, we slide the fresh ensemble mean into the 6-month input window, drop the oldest field, and repeat. This bootstraps the usable lead time from one month to roughly 24-30 months while keeping the computational cost modest.

\subsection{Datasets}

ENSO emerges from a labyrinth of ocean-atmosphere feedbacks in which zonal and meridional wind stress, subsurface temperature and salinity, and a host of other variables are interwoven. While each of these factors is indispensable to the genesis and evolution of ENSO events, SST remains the most conspicuous and immediate expression of the phenomenon~\cite{yeh_nino_2009,cai_increased_2022,li_insights_2025}. Guided by this physical primacy, we adopt SST as the sole input to the diffusion model, thereby isolating and evaluating the intrinsic ability of the model to capture the essential dynamics of the ENSO system.

In pretraining experiments using CMIP6 model data (1850-2014)~\cite{zhou_self-attentionbased_2023}, we used sea surface temperature anomalies (SSTA). To derive these anomalies, we removed long-term trends and climatological seasonal cycles from the monthly averaged fields. We confined the study region to 120$^\circ$E-90$^\circ$W and 20$^\circ$N--20$^\circ$S. Within this domain, we interpolated the SSTA data onto a regular grid with a $2^\circ$ zonal resolution and a $0.5^\circ$ meridional resolution (increased to $1^\circ$ between $5^\circ$S and $5^\circ$N). We assigned a value of zero to all land grid points and regions with missing data.
After preprocessing, we fine-tuned the model using the IAPv4 dataset (1940-1979), applying the same preprocessing steps to maintain consistency. We then evaluated the model's performance and generalization capability on IAPv4 data (1980-2021). Throughout the experiments, we normalized all data using MIN-MAX scaling to ensure standardization.

In experiments conducted without CMIP6 pretraining, we directly utilized SST data from the IAPv4 dataset~\cite{cheng_iapv4_2024}. Our analysis focused on the region spanning 160$^\circ$E to 80$^\circ$W and 20$^\circ$N to 20$^\circ$S, employing a uniform 1$^\circ$ resolution in both latitude and longitude. We assigned zero values to all land grid points and areas with missing data within this domain. Following these preprocessing steps, we trained the model using IAPv4 SST data (1940-1999) and subsequently evaluated its performance on the test period (2000-2023).

\section{Theoretical Analysis}\label{sec7}

Diffusion models differ sharply from conventional approaches that fit deterministic mappings between data points. They recast generation as a probabilistic process, learning the evolution of the data distribution step by step to capture intricate relationships~\cite{rombach_high-resolution_2022,yang_diffusion_2024}. To unpack the mechanisms behind their striking empirical gains, this paper addresses three questions.

\textbf{Can the reverse-time stochastic differential equation of a diffusion model faithfully capture the physics of the El Ni\~no-Southern Oscillation?}

Among nonlinear oscillators, the van der Pol model stands out for its ability to reproduce self-excited cycles and has been applied across physics and biology~\cite{chedjou_analog_2001,van_de_wouw_volterra_2002}. Within climate science, ENSO epitomizes such nonlinear dynamics: ocean and atmosphere lock into a tight, self-reinforcing loop that drives pronounced inter-annual variability. To illuminate the underlying oscilatory mechanism and sharpen forecasts, researchers recast ENSO as a recharge-discharge oscillator and distilled its evolution into a van der Pol form~\cite{jin_equatorial_1997,burgers_simplest_2005}:
\begin{equation}
	\begin{gathered}
		\frac{d^{2}T}{dt^{2}}+2\mu\left(\frac{T^{2}}{T_{c}^{2}}-1\right)\frac{dT}{dt}+\omega_{0}^{2}T=0, \\
		\mu=C-R_{h}, \\
		T_{c}^{2}=\frac{2\mu}{3\varepsilon}=\frac{C-R_{h}}{3\varepsilon}, \\
		\omega_{0}^{2}=DE-CR_{h},
	\end{gathered}
	\label{eq:van_der_pol_system}
\end{equation}

where $T$ denotes the sea-surface-temperature anomaly in the eastern equatorial Pacific, the positive parameters $C$, $D$, $E$, $R_h$, and $\varepsilon$ quantify, respectively, the intrinsic damping of SST anomalies, the strength of ocean-atmosphere heat exchange, the feedback from thermocline-depth anomalies onto SST, the natural relaxation rate of thermocline anomalies, and the nonlinear damping of the system. Together, these constants govern the oscillatory behaviour of ENSO's period, amplitude, and stability. Equation (\ref{eq:van_der_pol_system}) can be rearranged into an explicit relation between $dT$ and $dt$, giving Equation (\ref{eq:dT_equation}).
\begin{equation}
	dT = \left[ -\frac{T_c^2}{2\mu(T^2 - T_c^2)} \left( \frac{d^2 T}{dt^2} + \omega_0^2 T \right) \right] dt.
	\label{eq:dT_equation}
\end{equation}

\begin{figure}[htbp]    
	\centering            
	\includegraphics[width=0.7\linewidth]{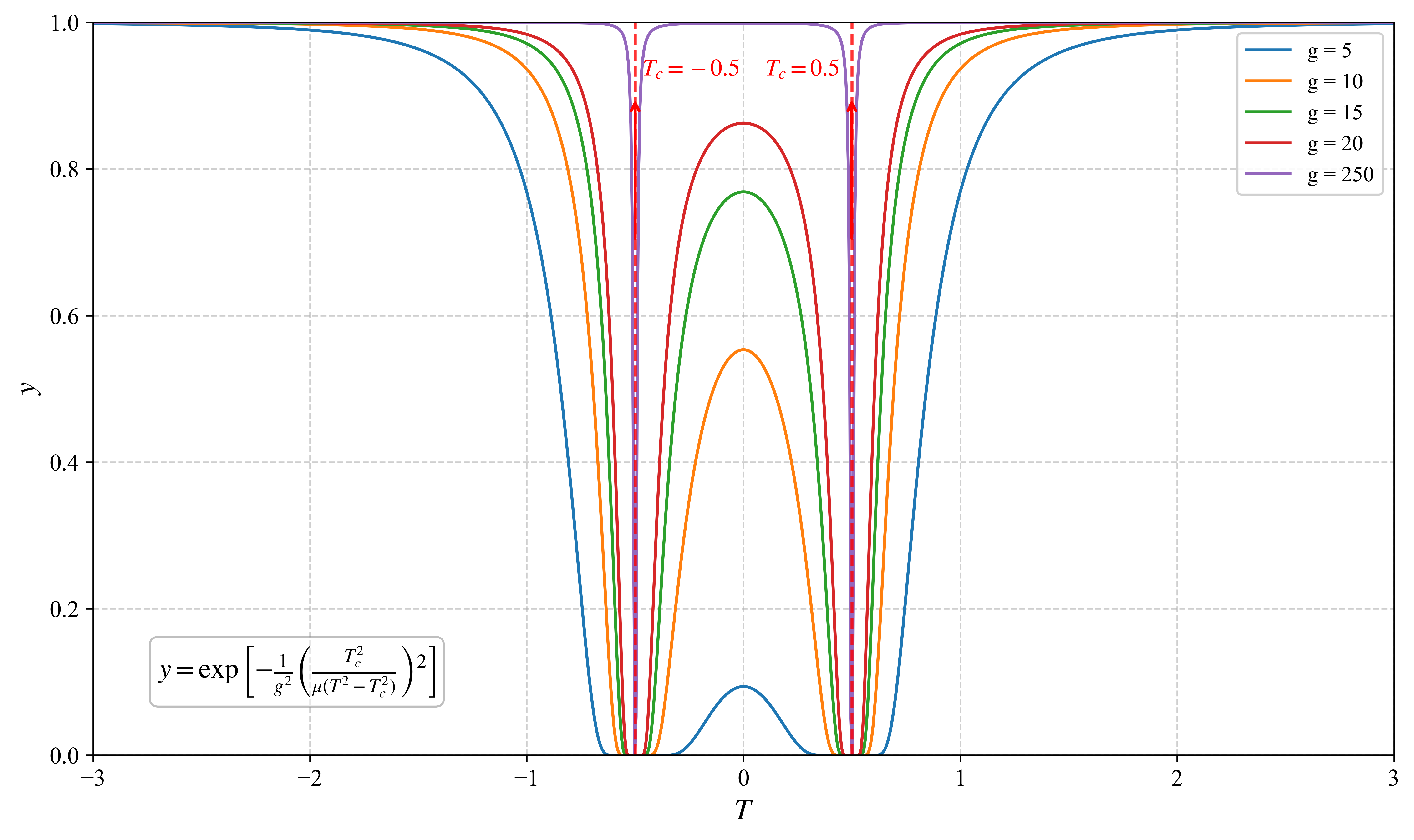}
	\caption{
		\textbf{Influence of parameter g on the assumed probability density.}}
	\label{fig:11}   
\end{figure}

Formally, the deterministic part of the reverse-time SDE in the diffusion model mirrors the van der Pol equation almost term-by-term. The two descriptions become equivalent once the following Equation (\ref{eq:system_cases}) hold.
\begin{equation}
	\begin{cases}
		f(T,t) = -\dfrac{T_c^2}{2\mu(T^2 - T_c^2)} \cdot \omega_0^2 T \\
		-\dfrac{1}{2} g(t)^2 \nabla_T \log P_t(T) = -\dfrac{T_c^2}{2\mu(T^2 - T_c^2)} \cdot \dfrac{d^2 T}{dt^2}
	\end{cases}
	\label{eq:system_cases}
\end{equation}

where $\nabla_T \log P_t(T)$ denotes the neural network's learned score-the gradient of the log marginal probability density-while $f(T,t)$ and $g(t)$ are hand-crafted functions. Physically, $\nabla_T \log P_t(T)$ points in the direction of steepest ascent of the probability density; the extra term therefore steers the sampling particle toward regions of higher density and, in doing so, reproduces the data distribution. Our focus is on this neural-network component. If we can identify a probability density whose score function matches Equation (\ref{eq:dT_equation}), the reverse process of the diffusion model will inherently encode the dynamics of a recharge-discharge van der Pol oscillator. We assume the probability density function given in Equation (\ref{eq:probability_distribution}). Fig.\ref{fig:11} visualises this density for several values of $g$, setting $\mu = 0.13$ and $T_c^2 = 0.25$.

\begin{equation}
	P_t(T) \propto \exp\left[ -\frac{1}{g^2(t)} \left( \frac{T_c^2}{\mu(T^2 - T_c^2)} \right)^2 \right],
	\label{eq:probability_distribution}
\end{equation}

where $T^2 \neq T_c^2$. In the model, $g(t)$ serves as the ``rate governor'' of the ENSO system. When $g^2(t)$ drops in boreal winter, the weight $g^-2(t)$ in the exponent rises, and the tail of $P_t(T)$ decays more steeply away from the equilibrium $T_c$. It matches observations: although ENSO events can be strong in winter, seasonal dynamics keep the system from lingering far from the critical state, so extreme departures remain relatively unlikely.
We interpret $T_c$ as the critical SST threshold. As $|T| \to T_c$, the denominator of the density approaches zero and the exponent explodes, forcing $P_t(T)$ to plunge toward zero. This sharp fall-off mirrors the nonlinear critical-region feedback documented in~\cite{boucharel_enso_2011,choi_enso_2013}: once the anomaly nears $T_c$, strong nonlinear feedbacks amplify any perturbation exponentially, rapidly expelling the system from the critical zone. Plugging the Equation (\ref{eq:probability_distribution}) into Equation (\ref{eq:system_cases}) yields
\begin{equation}
	\frac{d^2 T}{dt^2} = \frac{4}{\ln 10} \cdot \frac{T_c^2 T}{\mu (T^2 - T_c^2)^2}.
	\label{eq:second_derivative}
\end{equation}

We now classify the state of $T$ and qualitatively examine how Equation (\ref{eq:second_derivative}) behaves.

\textbf{Critical eruption } ($|T| \approx T_c$)
The denominator approaches zero, so the right-hand side of Equation (\ref{eq:second_derivative}) surges. This abrupt growth translates into a rapid increase in the acceleration of the SST anomaly-exactly the explosive departure from the quasi-steady state observed during the fast-growing phase of ENSO.

\textbf{Far from threshold} ($|T| \gg T_c$)
In this regime, the non-linearity weakens, and Equation (\ref{eq:second_derivative}) reduces to a second-order oscillator with mild damping. Substituting this weakly non-linear form into the ENSO van der Pol equation gives
\begin{equation}
	\frac{4}{\ln 10} \frac{T_c^2}{\mu T^3} + 2\mu \left( \frac{T^2}{T_c^2} - 1 \right) \frac{dT}{dt} + \omega_0^2 T = 0.
	\label{eq:modified_vanderpol}
\end{equation}
Because $|T| \gg T_c$, the term collapses to zero, yielding
\begin{equation}
	\frac{dT}{dt} = \frac{-\omega_0^2}{2\mu \left( \frac{T^2}{T_c^2} - 1 \right)} T
	\label{eq:dT_dt}.
\end{equation}
When $T > 0$, the system is in an El Ni\~no state and $dT/dt$ becomes negative; $T(t)$ then monotonically decreases, reflecting the physical fact that a strong warm anomaly damps back toward the climatological mean. The same argument holds for $T < 0$. This qualitative picture, together with Fig.\ref{fig:11}, supports the plausibility of the assumed density.
Yet Fig.\ref{fig:11} also reveals that the hypothesis cannot reproduce the observed statistical asymmetry between La Ni\~na and El Ni\~no. The underlying recharge-discharge theory itself lacks a mechanism for this asymmetry~\cite{an_enso_2018}, and the present assumption inherits that limitation. Future work can introduce an additional skew term to match observations. Similarly, the assumed function diverges when integrated over the full real line and cannot be normalized directly. Because proportionality only specifies the shape of the density, we truncate the temperature domain to enforce convergence and then apply data-driven normalization to obtain a valid probability density.
These two shortcomings-lack of asymmetry and the need for truncation are acceptable costs of a deliberately parsimonious model. They do not compromise the hypothesis's ability to capture the essential physics, and its practical value remains intact. Therefore, when the probability density satisfies Equations (\ref{eq:probability_distribution}) and (\ref{eq:second_derivative}), the reverse-time stochastic differential equation of the diffusion model subsumes the van der Pol equation of the recharge-discharge oscillator.

To establish a more general theoretical framework, we start with the governing equations of ENSO's recharge-discharge oscillator introduced by Jin (1997)~\cite{jin_equatorial_1997}. In this thermodynamic analogy, the La Ni\~na phase sees persistent easterlies pile up warm water in the western equatorial Pacific, charging the ocean heat reservoir. Once the heat content crosses a critical threshold, the warm pool surges eastward, triggering an El Ni\~no event that discharges the stored heat. Strengthened trades then drive the warm water westward again, resetting the cycle. The governing set of equations is:
\begin{equation}
	\begin{aligned}
		\frac{dT}{dt} &= CT + Dh - \varepsilon T^3, \\
		\frac{dh}{dt} &= -Et - R_h h,
	\end{aligned}
	\label{eq:system_equations}
\end{equation}
where $T$ denotes the sea-surface-temperature anomaly in the eastern equatorial Pacific, $h$ denotes the thermocline-depth anomaly in the western equatorial Pacific, and all coefficients in the equations are model parameters. For this study, we recast Equation (\ref{eq:system_equations}) into the reverse-time SDE form of the diffusion model, as shown in Equations (\ref{eq:dT_relation}) and (\ref{eq:dT_stochastic}).

\begin{equation}
	dT = \left( C T + D h(t) - \varepsilon T^3 \right) dt.
	\label{eq:dT_relation}
\end{equation}

\begin{equation}
	dT = \left[ f(T,t) - \frac{1}{2} g^2(t) \nabla_T \log q_t(T) \right] dt.
	\label{eq:dT_stochastic}
\end{equation}

The two equation sets are structurally almost identical. By carefully choosing the functional forms of $f(T,t)$ and $\nabla_T \log q_t(T)$, we can make them equivalent under specific conditions, which demonstrates that the reverse-time SDE of the diffusion model can indeed reproduce the ENSO dynamics. Concretely, we set
\begin{gather}
	f(T,t) = CT \\[6pt]
	\nabla_T \log q_t(T) \propto \exp\left[ -\frac{1}{g^2(t) \ln 10} \left( D h - \varepsilon T^3 \right)^3 \right].
	\label{eq:f_grad_logq}
\end{gather}

We find that once $T$ satisfies the specific condition given in Equation (\ref{eq:epsilon_equation}). at any time $t$, the models described by Equations (\ref{eq:dT_relation}) and (\ref{eq:dT_stochastic}) coincide exactly.
\begin{equation}
	9 \varepsilon^2 T^5 - 9 \varepsilon T^2 D h(t) = 2 \ln 10.
	\label{eq:epsilon_equation}
\end{equation}

In 2024, Jin and colleagues introduced the eXtended Recharge-Oscillator (XRO) model~\cite{zhao_explainable_2024}. By coupling the tropical Pacific with the Indian and Atlantic basins, XRO extends ENSO predictability to 16-18 months-well beyond the skill of conventional dynamical models. The key idea is to replace the scalar state variables of the classic recharge oscillator with a matrix that tracks interactions among multiple climate modes, while retaining nonlinear ocean-atmosphere feedbacks, seasonal modulation, and stochastic forcing. Crucially, the XRO still reduces to the same recharge-discharge dynamical core, so the reverse-time SDE of the diffusion model can subsume not only the original oscillator but also its modern XRO extension.

The diffusion model's reverse process provides a mathematically abstract yet physically meaningful representation of ENSO evolution. Through neural network learning of the required functions, we can reconstruct complete ENSO trajectories.
This physics-guided stochastic emulator achieves two key objectives. First, it probabilistically quantifies the intrinsic uncertainties in the climate system. Second, it effectively captures random environmental perturbations.
The approach yields dual benefits - enhanced understanding of ENSO dynamics and improved predictive capability. By implementing reverse-time stochastic differential equations, the method opens new possibilities for developing data-driven stochastic models of complex geophysical systems.

\textbf{Why can a simple U-Net capture the intricate nonlinear dynamics of the ENSO?} 

The key lies in how diffusion models decouple the historically intractable joint distribution between past and future system states. By unfolding this joint density into a succession of far simpler, conditionally Gaussian transitions, the model replaces the daunting task of learning a single, high-dimensional mapping with a sequence of low-dimensional, nearly linear sub-problems, each interpretable as a step in an evolving Gaussian mixture~\cite{shah_learning_2023,song_score-based_2020}. This staged decomposition concentrates the neural network, at every step, on the localized spatiotemporal structures that dominate that particular scale, so that it never has to confront the full complexity of the coupled ocean-atmosphere system all at once. Although the finer discretization increases computational cost, the benefit is decisive for ENSO. The network first distills the governing rules of small-scale perturbations and then, through hierarchical aggregation, fuses these local insights into a coherent large-scale portrait of the oscillation. Consequently, the layered learning strategy markedly enhances the model's ability to both represent and predict the pronounced nonlinearities that characterize ENSO.

The neural network is responsible for learning the high-dimensional mapping that links noisy states to the historical oceanic anomalies provided as conditioning information. For ENSO, whose nonlinearity emerges from long-range spatial coupling of sea-surface temperature anomalies, this objective meshes naturally with the strengths of a U-Net. The encoder progressively abstracts large-scale patterns, while the decoder, reinforced by skip connections, reconstructs fine-grained structures at key locations. Although Transformers surpass other architectures in modeling global dependencies, the U-Net captures localized spatial interactions at a markedly lower computational cost, and the diffusion process itself handles temporal evolution implicitly. Each denoising step discretizes the entire time axis, removing the need for an explicit long-range temporal module. Consequently, when the overall strategy relies on step-wise refinement, the encoder-decoder architecture and skip connections of a U-Net provide an efficient, purpose-built mechanism for spatial feature extraction in ENSO modeling~\cite{han_convective_2022,si_freeu_2024,chattopadhyay_towards_2022}.

In summary, the combined strategy of a diffusion-based stepwise simplification with U-Net spatial feature extraction proves highly effective for modeling ENSO-like phenomena whose evolution is governed by pronounced spatial organization. Rather than attempting to capture the entire complex dynamics in a single leap, this divide-and-conquer approach markedly reduces learning complexity while sharpening the representation of key spatial structures.

\textbf{Why don't diffusion models suffer the exponential error blow-up that plagues traditional deterministic autoregressive forecasts?}

Classical low-dimensional dynamical systems amplify even infinitesimal perturbations through positive Lyapunov exponents, so errors double (or worse) at a fixed rate along the forecast horizon~\cite{lorenz_deterministic_2004}. Diffusion models, by contrast, abandon the quest for a single best trajectory~\cite{price_probabilistic_2025,li_generative_2024,ho_denoising_2020}. Instead, they deliberately inject Gaussian noise at every step, reframing the forecasting problem as the evolution of a full probability distribution.
During the forward (noising) process, each diffusion step lifts the current state into a higher-dimensional phase space. A deterministic error that would otherwise concentrate along the temporal axis is scattered across countless noise dimensions. This dimensional expansion dilutes the error's local density, preventing it from compounding directionally in time. Consequently, the forecast never follows an ever-steepening exponential curve. Instead, the uncertainty spreads outward and remains bounded and controllable.

The essence of a diffusion model lies in learning the reverse-time probability flow. Instead of fitting a single deterministic state transition, it targets the entire conditional distribution $p(x_{t-1} | x_t)$. During autoregressive forecasting, the network denoises at every step given the historical context $y$: $x_{t-1} \sim p_\theta(x_{t-1} | x_t, y)$. Because the model is trained to recover the statistically optimal path from a noisy distribution rather than to chase exact numerical values, it shows strong robustness to small perturbations in $x_t$. Even if the preceding prediction drifts, the denoiser will pull it back onto a plausible trajectory as long as the drift stays within the typical set of the true data distribution. By combining noise-driven dimensional expansion with iterative error compression from the denoising network, the diffusion model converts the exponential error explosion of traditional schemes into a sub-linear accumulation. This transformation stabilizes long-range forecasts and is an effective tool for systems marked by intricate nonlinearities such as ENSO.

In aggregate, the superior performance of diffusion models in ENSO forecasting can be ascribed to the synergistic integration of physics-informed inductive biases, a stage-wise construction of the probabilistic flow, and an implicit error-regulation scheme. This confluence empowers the model to accurately replicate the spatio-temporal evolution of ENSO dynamics, faithfully characterize its aleatoric and epistemic uncertainties, and furnish long-horizon predictions whose error growth remains sub-exponential.

\backmatter

\bibliography{sn-bibliography}

\end{document}